\documentclass{elsart3p}

\usepackage{graphicx}

\usepackage{amssymb}

\begin{document}

\begin{frontmatter}



\title{Andreev bound states in rounded corners of $d$-wave superconductors}
\author{Christian Iniotakis}
\ead{iniotakis@itp.phys.ethz.ch}
\address{Institut f\"ur Theoretische Physik, Universit\"at T\"ubingen, Auf der Morgenstelle 14, D-72076 T\"ubingen, Germany}
\address{new address: Institut f\"ur Theoretische Physik, ETH-H\"onggerberg, CH-8093 Z\"urich, Switzerland}


\begin{abstract}
Andreev bound states at boundaries of $d$-wave superconductors are strongly
influenced by the boundary geometry itself. In this work, the zero-energy 
spectral weight of the local quasiparticle density of states is presented
for the case of  wedge-shaped boundaries with rounded corners. Generally, both orientation 
of the $d$-wave and the specific local reflection properties of the rounded
wedges determine, whether Andreev bound states exist or not. For the bisecting
line of the wedge being parallel to the nodal direction of the $d$-wave gap function,
strong zero-energy Andreev bound states are expected at the round part of the boundary.
  
\end{abstract}

\end{frontmatter}

\section{Introduction}
\label{}

At straight boundaries of  $d_{x^2-y^2}$-wave superconductors,
an enhancement of the zero-energy spectral weight can be found
in the local quasiparticle density of states \cite{Hu,Buchholtz,TK1}. This effect is due 
to Andreev bound states. And if the boundary is specular, their existence
strongly depends on the orientation of the boundary with respect to the $d$-wave gap function.
Experimentally, Andreev bound states can be observed as 
pronounced zero-bias conductance peaks in the tunneling spectrum \cite{TKRep,Deutscher}. 
Furthermore, locally resolved 
STM-measurements show a clear correlation between the tunneling spectra 
and the boundary geometry on the nano-scale \cite{Millo}. 

In a recent work \cite{CIPRB,CIDiss}, a variety of $d$-wave superconductors
with specular non-trivial boundary geometries
has been examined within the framework of quasiclassical theory \cite{Eilenberger,Larkin}. 
As a general result, apart from their orientational dependence, 
Andreev bound states are also strongly influenced by the shape 
of the boundary geometry on the length scale of
the coherence length $\xi$: Since the boundary geometry determines, how
incoming quasiparticles get reflected (and if they 'see' a 
sign change in the gap function), the local reflection behaviour 
of a non-trivial boundary geometry is the crucial factor for the 
existence of zero-energy Andreev bound states.

\section{Wedges and Andreev bound states}

\begin{figure}
\centering
\includegraphics[width=0.46 \columnwidth]{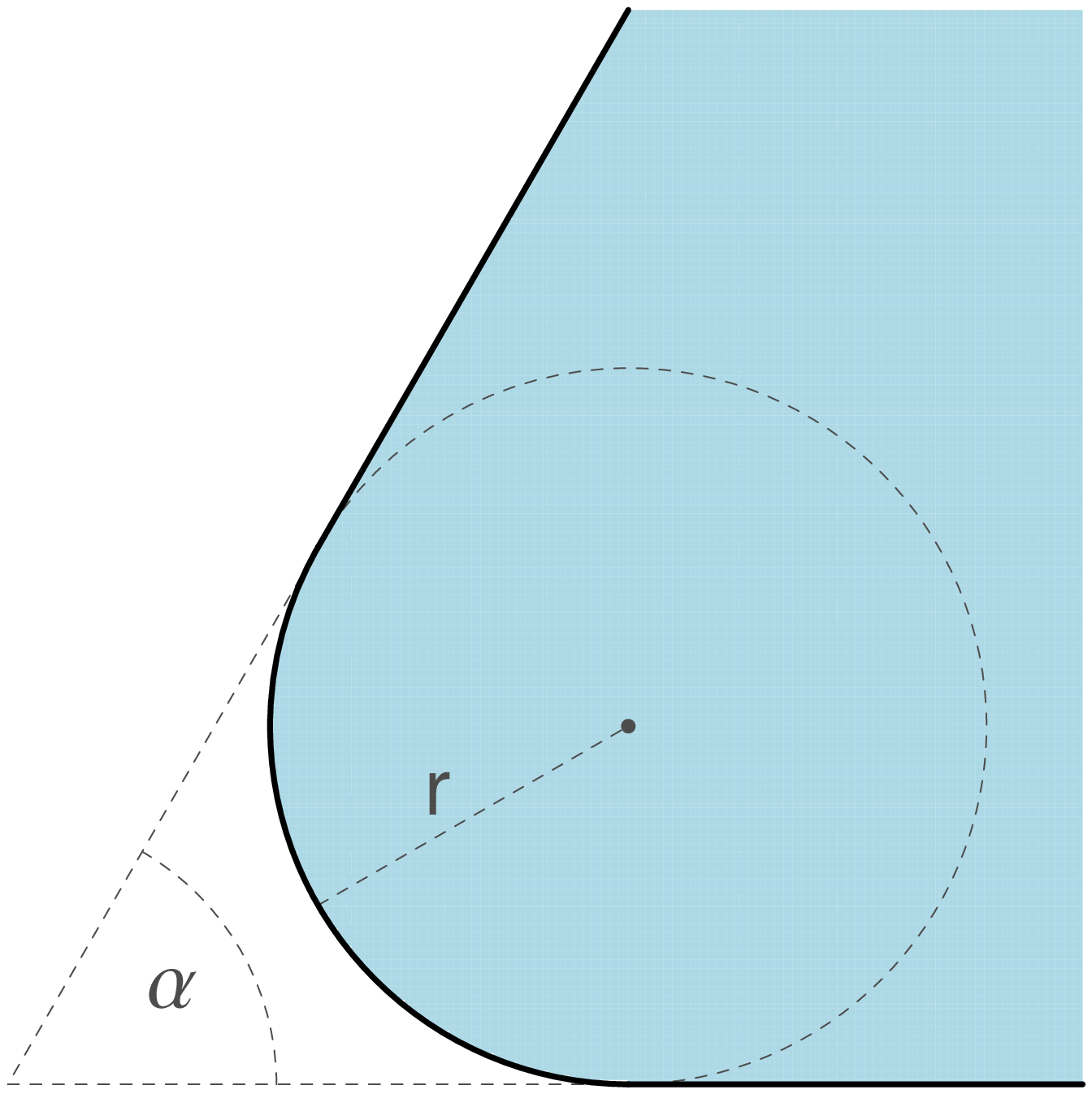}
\includegraphics[width=0.45 \columnwidth]{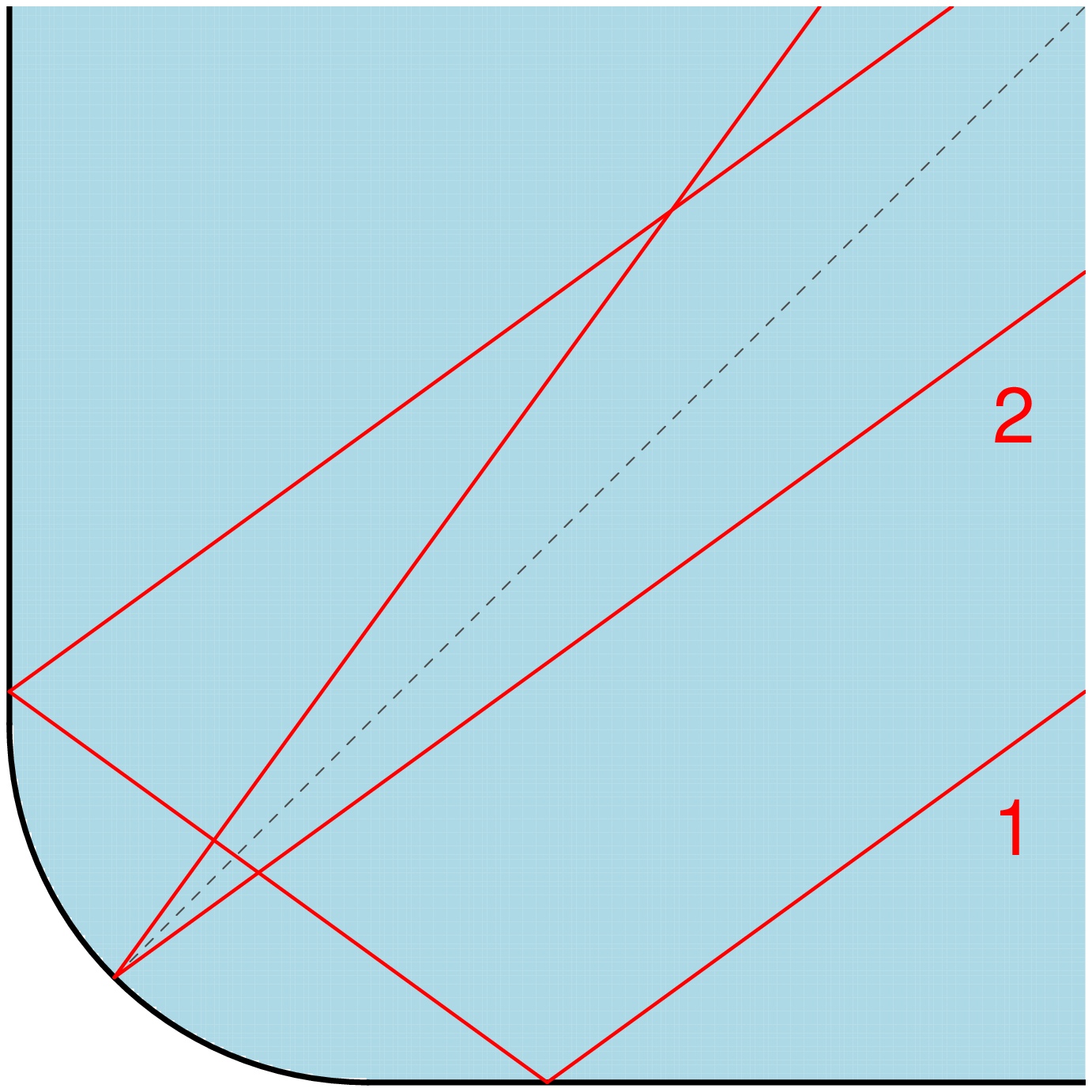}
\caption{\it(color online) \rm The boundary geometry of a rounded wedge is fixed by the opening 
angle $\alpha$ and the radius $r$ (left). The reflection properties of the wedge (1) and the round part  
of the boundary (2) may be completely different (right).}  
\label{Fig01}
\end{figure}

Wedges with sharp corners have different reflection properties,
depending on their opening angle. For opening angles 
$\alpha_n^+=\pi/(2n)$ incoming and outgoing quasiparticle 
trajectories are always parallel, whereas for opening angles 
$\alpha_n^-=\pi/(2n-1)$ the direction of the outgoing trajectory
is additionally mirrored at the bisecting line of the wedge.
If the bisecting line of the wedge itself is kept
parallel to the nodal direction of the $d$-wave gap function, this different
reflection behaviour has amazing consequences on the quasiparticle 
spectra in the corners, at least theoretically: 
The zero-energy spectral weight rapidly oscillates as a function
of the opening angle, with maxima appearing at the angles $\alpha_n^-$ 
due to Andreev bound states. Minima are situated at the opening angles 
$\alpha_n^+$, where zero-energy Andreev bound states do not exist \cite{CIPRB,CIDiss}.

\section{Wedges with rounded corners}

\begin{figure}
\centering
\includegraphics[width= \columnwidth]{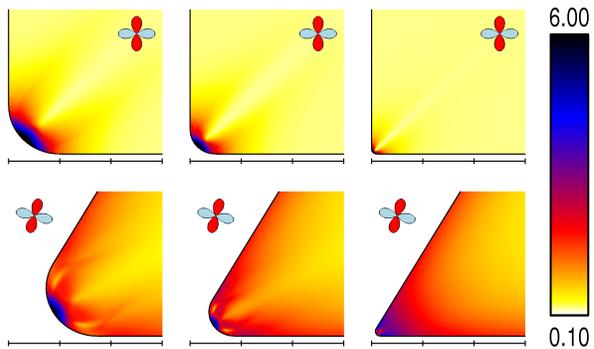}
\caption{\it(color online) \rm Zero-energy local density of states $N(E=0)$ in rounded
corners of $d$-wave superconductors. The bisecting line
of the wedges is parallel to the nodal direction of the $d$-wave, and the length
scale is in units of the coherence length $\xi$.
The opening angle of the wedges is $\alpha=\pi/2$ (upper row) and $\alpha=\pi/3$ (lower row).
The radius $r$ of the round part is (from left to right)  $r=\xi$, $r=0.5\xi$,  $r=0.1\xi$. }  
\label{Fig02}
\end{figure}

Wedges with rounded corners can be parametrized by an additional
radius $r$ determining the rounding of the boundary
geometry. However, the local reflection behaviour of the 
round part, which tends to mirror quasiparticle trajectories 
at the bisecting line of the wedge, may be different to that
of the original wedge, cf. Fig. \ref{Fig01}. Considering again the case,
 where the bisecting line of the wedge is parallel to the nodal
direction of the $d$-wave, this 'new' reflection type directly leads 
to zero-energy Andreev bound states close to the round part of 
the boundary geometry. A sharp right-angled
wedge with opening angle $\alpha=\alpha_1^+$, for example, 
does not exhibit zero-energy Andreev bound states, but they
are created by an additional rounding, as can be seen in Fig. \ref{Fig02}, 
upper row.

Similarly, strong bound states exist at the rounded corner of a
wedge with opening angle $\alpha=\pi/3=\alpha_1^-$, which is presented
in Fig. \ref{Fig02}, lower row. In contrast to the right-angled wedge,  
however, considerable zero-energy Andreev bound states remain in the corner region 
when the rounding is decreased, because they get induced by the wedge-shaped boundary 
geometry itself. Although the reflection properties of both rounded corner and wedge enhance
Andreev bound states separately, they are not identical (e.g. regarding the number of quasiparticle 
reflections). The competition between them may locally even lead to a destructive 
interference effect. 

\section{Acknowledgements}
The author thanks S. Graser, T. Dahm and N. Schopohl for valuable
discussions and is grateful to the German National Academic 
Foundation for financial support.

\end{document}